\documentclass[12pt]{article}
\usepackage{latexsym}  
\usepackage{amssymb}
\usepackage{graphicx}
\usepackage{amsmath}

\begin{document}

\begin{center}

{\Large\bf Another Formula for the Charged Lepton Masses }

\vspace{4mm}

{\bf Yoshio Koide}

 {\it Department of Physics, Osaka University, 
Toyonaka, Osaka 560-0043, Japan} \\
{\it E-mail address: koide@kuno-g.phys.sci.osaka-u.ac.jp}

\end{center}

\vspace{3mm}

\begin{quotation}
A charged lepton mass formula 
$(m_e +m_\mu + m_\tau)/(\sqrt{m_e}+\sqrt{m_\mu} + \sqrt{m_\tau})^2 
=2/3$ is well-known. 
Since we can, in general, have two relations for three quantities, 
we may also expect another relation for the charged lepton masses.  
Then, the relation will be expressed by a form of  
$\sqrt{m_e m_\mu m_\tau}/(\sqrt{m_e}+\sqrt{m_\mu} + \sqrt{m_\tau})^3$.  
According to this conjecture, a scalar potential model is speculated.
\end{quotation}

PCAC numbers:
  11.30.Hv, 
  12.60.-i, 


\vspace{5mm}


A charged lepton mass formula \cite{K-relation} 
$$
 K \equiv \frac{m_e +m_\mu + m_\tau}{(\sqrt{m_e}+\sqrt{m_\mu} + \sqrt{m_\tau})^2 }
= \frac{2}{3}
\eqno(1)
$$ 
is well-known. 
When we introduce U(3)-family nonet scalar $(\Phi)_i^{\ j}$ 
($i,j=1,2,3$) and we assume that the vacuum expectation value (VEV) 
on the diagonal basis of $\langle \Phi \rangle$
is given by 
$$
\langle \Phi \rangle = {\rm diag} (v_1, v_2, v_3) 
\propto  {\rm diag}( \sqrt{m_e} , \sqrt{m_\mu} , \sqrt{m_\tau}),
\eqno(2)
$$
the formula (1) is expressed as
$$
K= \frac{ [\Phi\Phi] } {[\Phi]^2} = \frac{2}{3}.
\eqno(3)
$$
Here and hereafter, for convenience, we drop the notations 
``$\langle$" and ``$\rangle$" in the VEV matrix 
$\langle \Phi \rangle$, and  we  denote Tr$[A]$ as $[A]$ simply.  

Since we can, in general, have two relations for three quantities, 
we may also expect another relation for the charged lepton masses.  
Since the relation (1) is invariant under a scale transformation
$(m_e, m_\mu, m_\tau) \rightarrow  (\lambda m_e, \lambda m_\mu, 
\lambda m_\tau)$, we speculate that the second relation will also 
be invariant under the scale transformation. 
Therefore, the relation will be expressed by a form of  
$$
\kappa \equiv \frac{ \sqrt{m_e m_\mu m_\tau} }{
(\sqrt{m_e}+\sqrt{m_\mu} + \sqrt{m_\tau})^3} 
= \frac{{\rm det} \Phi}{[\Phi]^3 } .
\eqno(4)   
$$
Since the relation (3) has been derived \cite{YK_MPLA90} from 
a scalar potential model based on a U(3) family symmetry,  
the value $\kappa$ in (4) will also be obtain a U(3)-family 
scalar potential model. 
According to this conjecture, a scalar potential model is speculated 
in this paper. 

First, in order to speculate the value of $\kappa$ defined by (4), 
let us give a brief review of the derivation of the value of $K$ 
given by (3), because the derivation of $\kappa$ will be done 
analogously to the derivation of $K$.   
In the Ref.\cite{YK_MPLA90}, the scalar potential $V$ has been 
given by 
$$
 V= \mu^2 [\Phi\Phi] + \lambda  [\Phi\Phi] [\Phi\Phi] 
 + \lambda'  [\Phi_8\Phi_8] [\Phi]^2 , 
\eqno(5) 
$$
where $\Phi_8$ is the octet part of the nonet scalar $\Phi$ 
defined by
$$
\Phi_8 \equiv  \Phi - \frac{1}{3} [\Phi] {\bf 1}, 
\eqno(6)
$$
where ${\bf 1}$ denotes a $3\times 3$ unit matrix.  
Note that the $\lambda'$-term is not U(3)$_{family}$ invariant, but 
still SU(3)$_{family}$ invariant. 
Since the derivative $\partial V/\partial \Phi$ leads to
$$
\frac{\partial V}{\partial \Phi} = 2 \left( \mu^2 + \lambda [\Phi\Phi] + 
\lambda' [\Phi]^2 \right) \Phi + 
2 \lambda' \left( [\Phi\Phi] -\frac{2}{3} [\Phi]^2 \right) [\Phi] \, {\bf 1} .
\eqno(7)
$$
Here, we have used $\partial [\Phi^n]/\partial \Phi = n \Phi^{n-1}$
and $\partial [\Phi]/\partial \Phi = {\bf 1}$.  
The requirement of $\partial V/\partial \Phi =0$ under the 
condition $\Phi \neq {\bf 1}$ leads to 
$$
\mu^2 + \lambda [\Phi\Phi] + \lambda' [\Phi]^2 =0,
\eqno(8)
$$
and
$$
[\Phi\Phi] -\frac{2}{3} [\Phi]^2 = 0 .
\eqno(9)
$$
The result (9) gives just the relation (3), and the result (8) fixes 
the scale of the VEV of $\Phi$. 
Note that the relation (9) has been derived independently of 
the potential parameters $\mu$, $\lambda$ and $\lambda'$. 
Therefore,  we may take the value of $\lambda'$ so that 
$\lambda'$ is negligibly small compared with the value of $\lambda$.  

Here, we should note that the ``charged lepton masses" 
$( \sqrt{m_e} , \sqrt{m_\mu} , \sqrt{m_\tau})$  in Eq.(2)  
are not pole masses (the observed masses), but 
mass values defined in the U(3)-family symmetry limit, 
as seen in the derivation (9).  
Although it is well known that the relation (1) is 
excellently satisfied with the observed charged lepton 
masses, this is nothing but an accidental coincidence.  
(However, Sumino \cite{Sumino_PLB09} has asserted  
that this excellent agreement is not accidental. 
In the present paper, we do not give a brief review of 
his article.) 


In this paper, stimulated by the success of the derivation (9)
from the potential (5), in order to derive another relation 
($\kappa$ value) for the ``charged lepton masses", 
we search a possible form of the potential  $V'$ which 
is another version corresponding to the third term 
($\lambda'$-term) in the potential (5). 
For such the purpose, we put the following guiding 
principles:  

\noindent [1] The potential $V'$ consists of U(3)$_{family}$ 
violated, but still SU(3)$_{family}$ invariant terms, i.e.
$[\Phi_8 \Phi_8 \Phi_8 \Phi_8]$, $[\Phi_8 \Phi_8 \Phi_8][\Phi]$,
$[\Phi_8 \Phi_8] [\Phi]^2$ and $[\Phi_8 \Phi_8][\Phi_8\Phi_8] $.

\noindent [II] $V'$ is given by a linear combination of such 
terms denoted in [I].  The coefficients among the terms 
must be given by simpler integers. 

According to the guiding principle [I], when we denote the general form 
of $V'$ as 
$$
V' = a_0 [\Phi_8 \Phi_8 \Phi_8 \Phi_8] + a_1 [\Phi_8 \Phi_8 \Phi_8][\Phi]
+ a_2 [\Phi_8 \Phi_8] [\Phi]^2 + a_4 [\Phi]^4 +  
a_{02} [\Phi_8 \Phi_8][\Phi_8\Phi_8] , 
\eqno(10)
$$
we obtain the relation
$$
\kappa = \frac{1} {27 a_1}  (4 a_0 -a_1 -72 a_4 +8 a_{02}) ,
\eqno(11)
$$
Note that the $a_2$ term does not contribute to the 
$\kappa$ relation as seen in (11). 
(In the potential (10),  the term $[\Phi]^4$ is  
a U(3)$_{family}$ invariant term against the guiding principle [I].  
Nevertheless, we added the $a_4$ term in addition to the U(3) 
violated terms $[\Phi_8 \cdots][\Phi]^n$ 
with $n= 0, 1, 2$. )

The relation (11) is obtained by using the following formulas: 
$$
 [\Phi_8 \Phi_8 \Phi_8 \Phi_8] =[\Phi \Phi \Phi \Phi] 
-\frac{4}{3} [\Phi \Phi \Phi ][ \Phi]
 +\frac{2}{3} [\Phi \Phi ][ \Phi ]^2 + \frac{1}{27} [\Phi]^4 ,
 \eqno(12)
$$
$$ 
 [\Phi_8 \Phi_8 \Phi_8][\Phi] = [\Phi \Phi \Phi ][ \Phi] 
- [\Phi \Phi ][ \Phi ]^2  +\frac{4}{9} [\Phi ]^4 ,
 \eqno(13)
$$
$$
 [\Phi_8 \Phi_8] [\Phi]^2 =[\Phi \Phi ][ \Phi ]^2 -\frac{1}{3}  [\Phi]^4 ,
 \eqno(14)
$$ 
$$
 [\Phi_8 \Phi_8][\Phi_8\Phi_8] = [\Phi \Phi][\Phi \Phi] 
- \frac{2}{3}  [\Phi \Phi ][ \Phi ]^2 +\frac{1}{9}   [\Phi]^4 .
 \eqno(15)
 $$
 Then, we obtain
 $$
 V' = a_0 [\Phi \Phi \Phi \Phi] 
 + \left( -\frac{4}{3} a_0 + a_1 \right)  [\Phi \Phi \Phi ][ \Phi]
 + \left( \frac{2}{3} a_0 -a_1 +a_2+ a_4 -\frac{2}{3} a_{02} \right) 
  [\Phi \Phi ][ \Phi ]^2   
 $$
$$ 
+ \left( -\frac{1}{9} a_0 + \frac{2}{9} a_1 -\frac{1}{3} a_2
+ \frac{1}{9} a_{02}   \right)  [\Phi]^4 
  + a_{02}  [\Phi \Phi ]  [\Phi \Phi ] , 
  \eqno(16)
  $$
  so that we obtain
  $$
  \frac{\partial V'}{\partial \Phi} = 4 a_0 \Phi \Phi \Phi 
  + \left( - 4 a_0 + a_1\right)  [\Phi]  \Phi  \Phi 
+ \left\{ \frac{2}{3} \left( 2 a_0 -3 a_1 -2 a_{02} \right) 
[ \Phi ]^2 + 4 a_{02}  [\Phi \Phi ] \right\} \Phi
$$
$$
+ \left\{ \left( -\frac{4}{3} a_0 +a_1\right)  [\Phi \Phi \Phi ] 
+ \frac{2}{3} \left( 2 a_0 -3 a_1 +3 a_2 -2 a_{02} \right) 
 [\Phi \Phi ][ \Phi] \right.  
 $$
 $$
 \left. 
 +\frac{4}{9} \left(- a_0 + 2 a_1 -3 a_2 + 3 a_4 +  a_{02} \right) 
  [\Phi]^3 \right\} {\bf 1} .
\eqno(17)
$$ 
Finally, we use  formulas for an arbitrary  $3\times 3$ 
Hermitian matrix $A$: $AAA= [A] AA +\frac{1}{2} ([AA]
-[A]^2) A + {\rm det} A {\bf 1}$ and $[AAA] = 3 {\rm det} A
+\frac{3}{2} [AA] [A] -\frac{1}{2} [A]^3$. 
Then, we obtain the relation (11) from the condition 
that the coefficient of the matrix ${\bf 1}$ must be zero. 
(In the result (11), we have used the relation (9), so that 
the factor $[\Phi\Phi]$ has been replaced  into 
$(2/3)[\Phi]^2$.) 
Here, we did not show coefficients of the matrices  
$\Phi\Phi$ and $\Phi$, because 
 those coefficients contain additional potential parameters 
$\mu$ and so on, as seen in the constraint (8), 
 so that we cannot obtain any meaningful constraint 
from those coefficients. 

As seen in Eq.(11), we cannot give a reasonable value of $\kappa$ 
as far as we choose only one term in (10), differently in the 
derivation (3). 
There are many combinations which satisfy the condition (11). 

Therefore, we use the second guiding principle [II].  
According to  [II],  we choose the following  simple form
$$
V' = \lambda' \left(  [\Phi_8 \Phi_8 \Phi_8 \Phi_8] + 
[\Phi_8 \Phi_8 \Phi_8][\Phi] + [\Phi_8 \Phi_8] [\Phi]^2 + 
\frac{1}{3^4} [\Phi]^4 - \frac{1}{4}   
 [\Phi_8 \Phi_8][\Phi_8\Phi_8] \right) . 
\eqno(18)
$$
Here, we took the coefficient $a_4$  as 
$a_4 =1/3^4$ correspondingly to the term 
$[\Phi_8 \Phi_8 \Phi_8 \Phi_8]$ with a replacement 
$\Phi_8 \rightarrow \frac{1}{3} [\Phi]$. 
For $a_{02}$,  we took $a_{02} = -1/4$. 
The form (18) is likely under the criterion ``simple",  
but the choice does not have theoretical basis. 

The potential form (18) can give an interesting prediction 
of $\kappa$  
$$
\kappa = \frac{1}{18 \times 27}  =\frac{1}{486} ,.
\eqno(19)
$$
independently of the coefficient $\lambda'$ in (18). 
On the other hand, the  observed value of $\kappa$  
 is given by 
$$
\kappa^{obs} = \frac{1}{486.663} , 
\eqno(20)
$$
where we have used the pole masses of the charged leptons 
\cite{PDG}. 
By recalling that our parameters $(m_e, m_\mu, m_\tau)$
do not mean the observed charged lepton masses,  
we think that the value (19) is reasonable.


In conclusion, we find a simple form of the potential $V'$ 
which can give a reasonable value of $\kappa$. 
However, since the form (18) is not a unique solution 
from the theoretical point of view, 
it is possible that there is another suitable form of $V'$. 
It is not essential whether the prediction excellently agrees 
or not with the observation. 
The noticeable point is what this scenario suggests to us.   
We notice a fact that the two similar scenarios lead to the two 
independent relations (1) and (19) for three values 
$(m_e, m_\mu, m_\tau)$, independently of the value of $\lambda'$.  
Therefore, the scenario seems to suggest the following picture:   

\noindent (i) Charged lepton masses are described 
in terms of bilinear form. 

\noindent (ii) U(3) family symmetry is worthy to 
notice.

\noindent (iii) Charged lepton mass 
spectrum  (Yukawa coupling constant) originates  
in VEV of a U(3) family nonet scalar. 
Therefore, there is a possibility that the spectrum 
can be understood from a form of the scalar 
potential. 

It will urgently become important to study 
a role of U(3) family nonet scalar (and also family 
gauge bosons) experimentally and theoretically.  
This is a future task to us.  
 

This work was supported by JSPS KAKENHI Grant No. JP16K05325.  

\vspace{5mm} 

{

\end{document}